# Exploring software developers' work practices: Task differences, participation, engagement, and speed of task resolution

Sherlock A. Licorish[✉], Stephen G. MacDonell

*Department of Information Science, University of Otago, PO Box 56, Dunedin 9054, New Zealand*
sherlock.licorish@otago.ac.nz, stephen.macdonell@otago.ac.nz

**Abstract**

*In seeking to understand the processes enacted during software development, an increasing number of studies have mined software repositories. In particular, studies have endeavored to show how teams resolve software defects. Although much of this work has been useful, we contend that large-scale examinations across the range of activities that are commonly performed, beyond defect-related issues alone, would help us to more fully understand the reasons why defects occur as well as their consequences. More generally, these explorations would reveal how team processes occur during all software development efforts. We thus extend such studies by investigating how software practitioners work while undertaking the range of software tasks that are typically performed. Multiple forms of analyses of a longitudinal case study reveal that software practitioners were mostly involved in fixing defects, and that their engagement covaried depending on the nature of the work they were performing. Furthermore, multiple external factors affected speed of task resolution. Our outcomes suggest that behavioral and intrinsic issues may interact with extrinsic factors becoming significant predictors of the speed of software task resolution.*

**Keywords:** Software developers, Task differences, Participation and engagement, Speed of task resolution, Mining software repositories, Empirical studies

## 1. INTRODUCTION

Software development is generally acknowledged as an intellectually challenging activity, and one that typically requires team members to work collectively to create a product or service that may be critical but yet is conceptual, fluid, and intangible. In spite of the ongoing provision of innovations with respect to development methods and tools [1–3], research continues to note that many software projects do not succeed [4–6]. In recent years, an increasing proportion of research effort has therefore aimed to understand the work practices and team processes implemented during development, in the belief that this knowledge might be better leveraged to improve project outcomes [7,8]. In particular, automatically generated archives and repositories have gained prominence as sources of information for those studying team behaviors, enabling researchers to study software practitioners' involvement in detail, and performance in development and maintenance activities. Such studies have examined both open and closed repositories of projects including Apache [9], Eclipse [10], GNOME, NetBeans, and OpenOffice [11], along with Jazz [12–15] and Windows Vista [16], providing explanations for various team phenomena and development issues and outcomes.

This growing attention given to mining software repositories reflects the emerging significance of the study of automatically stored software artifacts in order to understand team processes. More specifically, communication artifacts such as electronic messages, change request histories, and blogs are able to provide unique perspectives on the activities that occur during the software development life cycle (SDLC) – perspectives that cannot be drawn from code or similar technical artifacts. Analysis opportunities presented by these automatically recorded artifacts are also potentially valuable because of the reduction of the likely bias that can arise with self-reporting [17,18], as well as the unobtrusive nature of the investigation of team processes and work practices from such artifacts.

Although previous work has provided useful insights into the nature of the relationship between the frequency of practitioners' communications and the prevalence of bugs (or software defects) [10,19], comparatively less attention has been given to the ways in which developers work and behave when undertaking other software development tasks that are commonly performed [13,20]. In addition, although previous work has examined how communication patterns relate to team size [21,22], there is growing support for the need to examine the details within developers' communications beyond assessments based only on message exchange frequency [23,24]. This need to move beyond frequency- based assessments and to examine a larger spread of software tasks is supported by outcomes from early works in the social and organizational psychology space, which established that multiple properties of team tasks affect team performance [25–27]. Thus, different task ecosystems are likely to benefit from particular starting configurations and team arrangements. Exploring the way developers work across the range of software tasks that are commonly performed, and particularly those initial events that lead to the development of software features in the first place, and then subsequent bugs, could provide added value for the software development community.

We set out to address this research opportunity, and hence provide initial insights into the way software tasks are distributed in a large software project, and how teams assemble and undertake different forms of software development activity. This study thus provides an extension to those mentioned above, with a view to explaining how different task ecosystems are likely to benefit from particular team configurations. Our contributions in this paper are twofold: 1. We extract and mine a large software repository and apply both statistical and deeper content analysis (CA) approaches in our study of software artifacts to understand developers' work practices when performing various forms of software task. Our experiences are systematically documented, and may serve as a guide for those undertaking similar studies involving multiple forms of analyses. 2. Subsequently, given our observations, we decompose our findings through a range of theoretical lenses, provide multiple recommendations for software project governance, and identify avenues for future work.

The remainder of this paper is organized as follows. In the next section (Section 2), we survey previous work, and outline our specific research questions (RQs). We then describe our research model, and further decompose our RQs in Section 3. We explain our research method and measures in Section 4, also outlining our study context in this section. In Section 5, we present our results, and these findings are discussed in Section 6. In Section 7, we highlight the implications of our findings, and outline future research directions. In Section 8, we consider our study's limitations, and finally, we draw conclusions in Section 9.

## 2. BACKGROUND AND MOTIVATION

We survey previous work in this section. In order to both ground and structure our literature inspection, we first review those studies that have considered software teams' communication in Section 2.1. Considering our objective to examine how practitioners work across a range of software development tasks, we next examine general works on task differences in Section 2.2. We finally survey works that have considered software development task differences in Section 2.3, identifying research gaps and outlining our RQs in this subsection.

### 2.1. The study of software teams' communication

Beyond using and interpreting measurements focused on code and numbers of bugs [28,29], the availability of publicly searchable communication artifacts has provided researchers with the opportunity to study patterns in software development in far greater detail than would be possible if they were to consider technical artifacts alone [10,21]. This form of evidence has supported analysis at individual and team levels, providing insights in relation to participants' contributions to code. Such outcomes are evident in the literature on the study of communication and coordination from both open-source software (OSS) and closed-source software (CSS) repositories. In the OSS context, for instance, Abreu and Premraj [10] examined the Eclipse mailing list and found that developers communicated most frequently at release and integration time, but that increased communication also coincided with a higher number of bugs being introduced. Bird et al. [9] confirmed that the more software development an individual undertakes, the more coordination and controlling activities (s) he must perform. In considering patterns of contribution, Cataldo et al. [21] found that the practitioners who communicated the most also contributed most actively during software development. Furthermore, in a later study, Shihab et al. [30] found that proposals and actions discussed during team communication correlated with subsequent software development actions that were enacted, when studying the GNOME OSS project.

Earlier work by Howison et al. [22] found that a few key members of smaller OSS projects occupied the center of their teams' communication networks, in contrast to larger teams whose communication networks appeared more modular. Hinds and McGrath [31] confirmed the centralized communication pattern, but did not consider team size. Bird et al. [32] examined communities and subcommunities among the Apache, Python, PostgreSQL, and Perl projects, and concluded that specific technical needs drove the emergence of subcommunities in these projects. The Debian mailing list was interrogated by Sowe et al. [33] to observe knowledge-sharing among developers; they found that no specific individual dominated knowledge-sharing activities. A study of coordination conducted by Ehrlich et al. [34] found that brokers bridge communication gaps for teams that communicated across distributed sites. In addition, Ghapanchi [35] found that practitioners' willingness to communicate in OSS projects was positively influenced by task identification, and the popularity of such projects is positively influenced by who gets assigned tasks, and how such tasks are managed. These findings indeed support the view that evidence drawn from communication and coordination processes could reliably complement and support code-centric analyses, thus highlighting the importance of studying communication artifacts.

Beyond those works using OSS repository data to investigate team processes, artifacts in the IBM Rational Jazz CSS repository have also been used to study software practitioners' interactions and communications, largely from a social network analysis (SNA) perspective [36–38], offering somewhat contradictory findings to those noted in the OSS body of work. Contrary to the findings reported by others, which showed that a few developers generally dominate team communication [30,39], Nguyen et al. [37] studied multiple software teams at IBM Rational and noted that a high proportion – approximately 75% – of IBM Rational Jazz's team members actively participated in the project's communication network. In addition, these authors found IBM Rational Jazz project teams to have highly interconnected social networks, requiring few brokers to bridge communication gaps. When examining Jazz teams, and others developing software for the automotive industry, Ehrlich and Cataldo [40] found that there were improvements in teams' productivity and product quality when technical leaders shared more information, or when they occupied central positions in communication networks.

The derivation of insights such as those just described further supports the relevance of studying teams' communications. In fact, Datta et al.'s [41] SNA study of agile developers' collaboration while using the IBM Rational Jazz platform found that developers' expressions during regular communication possessed a wealth of useful information, much more than could be gleaned from examining source file changes alone. An earlier study (published in 1994) exploring software developers' activities found that up to 50% of practitioners' time was spent on interpersonal communication and coordination during software problem solving [42].

Studying these interactions therefore has the potential to reveal the reasons for, and consequences of, communication and coordination actions during software development projects. In this study, we use relevant (noncode) software artifacts to uncover such insights. In particular, we consider how software developers interacted and performed while undertaking different forms of software development task, an area that has attracted limited research to date. Accordingly, we next provide grounding for our study through an examination of the frameworks/models around task differences (Section 2.2).

**2.2. Task difference frameworks/models**
The work of McGrath and colleagues examining groups' and individuals' interactions and performance questioned the validity of outcomes derived from studies that failed to consider the characteristics of the tasks being undertaken [26,27]. They claimed that the nature of the work in which teams are involved underpins the differences that are implicit in team task performance. Outcomes will likely be optimized, for instance, if those who are naturally socially inclined perform tasks that may be deemed socially driven, while those who are motivated by intellectual work are drawn to conduct intellectual tasks. More generally, aligning an individual to a particular form of task given their specific orientation would naturally help with task performance [13,43,44].

A more granular classification of team tasks had been previously introduced by Carter et al. [45]. These authors classified team tasks into six types: clerical, discussion, intellectual construction, mechanical assembly, motor coordination, and reasoning. Similar to McGrath's analysis [26], Carter et al. [45] contend that each class of task has different performance processes, and hence, their successful completion demands different types of team behaviors and arrangements. In fact, in their empirical study of team leadership behaviors for facilitating various forms of tasks, Carter et al. [46] found that, while there were some general behaviors exhibited by all leaders, specific tasks indeed demanded some particular behaviors from leaders in order to succeed. A subsequent review of small group research conducted by McGrath and Altman in the 1960s [27] also provided a similar classification to that of Carter et al. [45], with an emphasis on the relevance and necessity of certain behaviors for specific tasks, thereby validating Carter et al.'s earlier position [45].

Subsequent work by Shaw [47] extracted six dimensions of group tasks when examining previous work: intellective versus manipulative requirements, task difficulty, intrinsic interest, population familiarity, solution multiplicity versus specificity, and cooperation requirement. Although the first of Shaw's dimensions refers to the property of a particular task, others consider the relationship between the task and those undertaking the task (i.e., the second, third, and fourth dimensions), how the task is evaluated (the fifth dimension), and how individuals must behave while working together to achieve the task outcome (the sixth dimension). Beyond Shaw's broad classification, the task model of Hackman [25] and Hackman and Morris [48] focused on intellectual tasks that led to written products. Their research revealed that three forms of task led to the delivery of such products: production tasks (tasks requiring idea generation or creativity), discussion tasks (tasks requiring dialog), and problem-solving tasks (tasks involving the execution of a plan).

In other related works by Steiner [49,50] classified tasks as either divisible or unitary. Divisible tasks are performed by many members of a group who are each responsible for specific task outputs, and where the performance of the team is equivalent to the combined performances of its members. However, unitary task outcomes are derived from the contributions of the team, and this may be disjunctive (the best member's ability may be considered equivalent to that of the group's), conjunctive (the team's task performance is equivalent to the weakest member's performance), or additive (task performance is equivalent to the average group performance). Steiner also posited that task performance may be influenced by motivation and coordination losses (or gains). The relevance of studying task differences has been further validated by others, with additional classifications offered by Laughlin and his colleagues [51,52]. These studies and models all contribute credence to the necessity of understanding and accounting for task differences when evaluating teams' performance.

Although several studies have considered the way individuals work from one team context to another [20,41,53–59], research examining software development teams' performance has given limited consideration to the study of individuals' and teams' engagement, behaviors and performance when undertaking a range of software development tasks. In fact, although there have been calls to consider the nature and structure of software tasks when considering software teams' communications and performance [60,61], little effort has been directed to this cause [13,20]. We address this gap in this study. In order to position our work in this context, the next subsection (Section 2.3) surveys previous work that has sought to examine software development task differences, with a view to elaborating our specific RQs.

**2.3. Software development task differences**
As noted in the preceding subsection, while the relevance of task differences to task performance has been previously established, and the aforementioned classification schemes have been considered by some previous research (e.g., [62,63]), little effort has been directed to studying variances in teams' engagements, attitudes, and performance while undertaking different forms of software development task.

In addition, while the above review illustrates that there has been a growing body of evidence aimed at providing understandings and explanations around software teams' communications and the social processes involved in software development (e.g., see [10,39]), limited previous research has considered how task difference frameworks/models may apply in the software development context [13,20,60,61]. Opportunities therefore exist for researchers to explore this issue in depth in order to enhance software delivery performance [4,64]. Having embraced the substantial body of knowledge around group work, task structure, and team performance [25,27,45,48,52], social and organizational psychology literature has been previously recommended as a relevant theoretical basis on which to underpin the study of software teams' processes. The basis for such recommendations is that, in addition to team composition and team norms, software team output is indeed, which is also affected by task structure [65]. In software development settings, it is considered that specific task ecosystems may influence practitioners' work ethics and levels of engagement, which may in turn affect team performance [66]. Although other researchers have provided a degree of anecdotal support for this proposition [60,61], and we ourselves have also endeavored to shed light on this phenomenon [20], research aimed at understanding software teams (and particularly those mining software repositories) has largely ignored the opportunity to provide firm empirical evidence around this subject.

Instead, most of the research considering differences in teams' engagement and performance patterns around software tasks has been focused on modeling the consequential incidence of bugs [67–72]. Although research efforts aimed at classifying and predicting bugs are indeed likely to be relevant and useful for reducing the incidence of defects, it is our contention that understanding the way in which developers work across the range of software tasks, and particularly those initial conditions that lead to the development of software features in the first place, and then the subsequent introduction of bugs, may provide added value for the software development community. Indeed, such efforts should be incremental, first aimed at understanding and explaining the complexities of team members' engagement and performance across the range of tasks that are regularly performed before, then exploring possible behavioral and competency requirements for undertaking different forms of software tasks, and examining the variance (or lack thereof) of members' attitudes on task performance. In line with this incremental stance, we therefore outline our initial RQ, aimed at examining task distribution and team members' participation around a large volume of software tasks in the first instance:

**RQ1.** (a) How are software tasks distributed in a large software project, and (b) how do team members participate in the different ecosystems of tasks?

Although software development as a whole is generally accepted as being complex due (in part) to the nature of the software product, which can be seen as conceptual and intangible in nature [73], all software tasks are not equal. For instance, coding a new feature or undertaking feature enhancement is likely to necessitate extensive intellectual and cognitive processes [45]. Such tasks will also present a different level of difficulty, and may require different team configurations and idea generation pro- cesses [74], when compared with tasks related to documentation, design, or software support. These latter tasks are likely to demand higher levels of manipulative and cooperative requirements, where social processes may feature more prominently, particularly during stakeholder consultation [47]. This need for higher levels of cooperation when undertaking requirements gathering and software design has indeed been observed by those examining the collaboration patterns of software developers [75]. In addition, increased consensus [76] is likely to benefit those working on documentation and software support and design tasks. Similarly, individuals resolving defects will require familiarity and specific problem-solving knowledge of the previously developed feature, and hence, such activities may demand less cooperation, tending toward smaller groups and increased individual focus and intellectual processes [25].

Aligned with such assessments, and because of perceived differences in software task purpose and focus, those working in the personality psychology and human resource management domains have stressed that team roles and software development positions should be supported by behavioral (and competency) profiling [77–80]. Under such an approach, specific tasks would be assigned to those occupying individual software roles; for example, software testers generally validate or verify that software features work as intended and are bug-free, while software analysts typically decompose the software application domain and prepare requirements for developers. Thus, in some ways, recommendations related to roles and positions, even now, indirectly affect software task assignments. As noted above, we have also in our own previous work sought to provide tangible recommendations, given preliminary findings from a small-scale study, as to how individuals should be selected given a specific portfolio of software tasks [13,20].

In spite of the studies just described, there remains a need for research to provide more conclusive evidence in terms of how developers engage across the range of software tasks that are commonly undertaken. This is particularly necessary given the evidence provided in the literature on the relevance and importance of task differences [25,27,45,48,52], along with our preliminary findings that suggest that various team and individual competencies may indeed be useful for performing particular tasks [20]. Accordingly, in line with frameworks/models considering task differences [25,27,45,48,52], we contend that different task ecosystems may require specific team participation and engagement requirements [7,77–80] – the absence of which may have implications for team performance. In fact, such requirements may also extend to the behavior and temperament of individual practitioners within teams or subgroups [2,13]. Such a research agenda has the potential to add to the body of knowledge in respect to such theories. In addition, grounded understandings of the ways in which software practitioners operate under various task configurations should also enable us to provide practice-based recommendations on how to plan for and arrange appropriate staffing to address different portfolios of software tasks. Such knowledge would be especially beneficial to the software development community if it is derived from self-organizing teams that have succeeded in delivering high-quality software products.

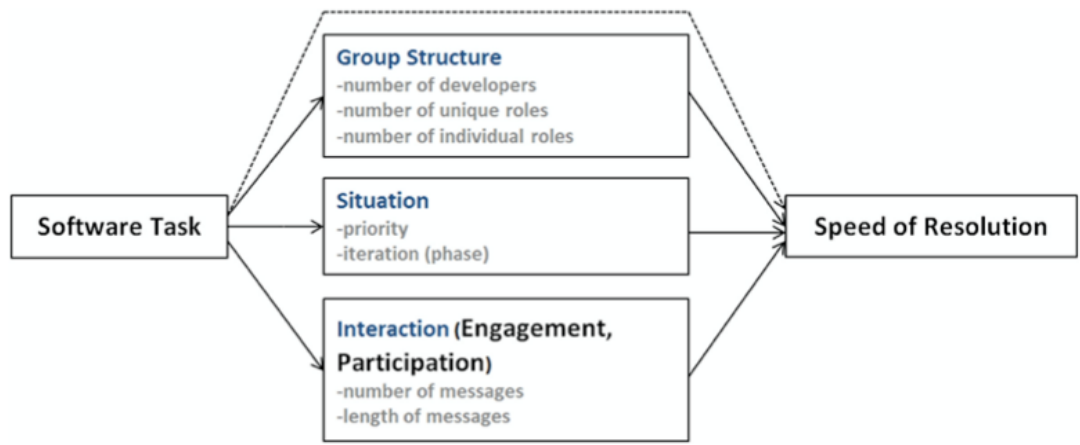

Fig. 1. Initial conceptual model of the relationships among software tasks, mediators, and speed of resolution.

Given these requirements and expectations, and leveraging theoretical support from those who have previously examined group work and task differences [25–27], we address two further RQs. Specifically, we examine artifacts from distributed teams that have successfully delivered commercially usable software to contribute insights into the way software teams' engagement covaries while undertaking various forms of software development tasks:

**RQ2.** Do the level and type of team members' engagement covary with the nature of the task they are performing?

**RQ3.** (a) Are there variances in speed of resolution for different forms of software task, and (b) what factors affect the speed of resolution of software tasks?

## 3. CONCEPTUAL MODEL

In the present study, we examine the way software practitioners participate in and engage around software tasks, as well as how quickly they resolve those tasks. We use small-group research and task difference theories as our theoretical bases – these emphasize the need to consider aspects of the task being undertaken when examining a team's performance [25–27]. We argue that understanding how individuals and teams participate and engage when undertaking various forms of software task would provide multiple outcomes for both research and practice (refer to Section 2.3).

Group interaction and task difference theories [25,27,45,48,52] support our initial conceptual framework, shown in Fig. 1, which illustrates the way software tasks possibly influence practitioners' participation, engagement, and speed of task resolution. The various components of the model in Fig. 1 align with RQs RQ1, RQ2, and RQ3 introduced above. We initially explore the differences in software activities and practitioners' participation and engagement in order to study differences in teams' interactions when they are working on various forms of software tasks (RQ1 and RQ2). Specific ecosystems of tasks may influence practitioners' levels of engagement and participation [66], which may in turn be affected by how much information is exchanged, those who are involved in these exchanges, when such exchanges take place, and the urgency with which action is to be taken [26]. Given the general challenges and complexities associated with measuring psychological constructs, such variables may also interact during members' participation and engagement [26,27], though we have focused less on these interactions in this work given our current motivation and initial exploratory stance. These interactions, and those mentioned above, are depicted in Fig. 1 in support of our analyses to answer RQ1 and RQ2 (refer to Section 4.2 for further details).

Answers to the first two RQs (RQ1 and RQ2) will provide the basis for the subsequent enquiry aimed at exploring the speed of resolution of various software tasks (RQ3). As multiple factors (including those related to group structure, the situation under which the task is being performed, and team members' interactions) have been shown to mediate teams' performance on group tasks [26,27], Fig. 1 also incorporates these factors as potential mediators in our conceptual framework. In Fig. 1, a number of subfactors are to be considered under each of these main factors (i.e., Group Structure, Situation, and Interaction), with those considered in the study of participation and engagement also being of relevance when examining speed of resolution. Previous work has shown that various factors affect the speed of task resolution, including the task itself [26,27]. Beyond the task, however, group properties such as the number of members and their roles [81], the urgency attached to the task and when it is scheduled [37], and who communicates and how frequently they do so [82,83], all affect the pace at which tasks are performed.

Hackman previously linked group performance to task activity and group structure [84]. Furthermore, Senior [85] established that the selection of group roles and structure is also influenced by the task that is being performed. In fact, specific arrangements of members will be configured given the needs of the tasks under consideration [85], and particularly in a software development context. For instance, during software development, some forms of defects or bugs may threaten product acceptance by the client, and hence, specific grouping(s) of individuals and roles may be assembled to fix these on the expectation of a high chance of success.

Similarly, the nature of the task teams and individuals assemble to perform may affect the situation under which these are resolved and the priorities that are attached to these tasks. Cottrell et al.'s [86] evaluation and apprehension theory shows that individuals' and teams' actions are influenced by their perception. If specific software tasks are assessed as critical to a team's performance evaluation, then members (and leaders in particular) are likely to increase their urgency. Thus, various software tasks may also influence the situation and configuration under which teams assemble to resolve them. All these variables will in turn influence the speed with which tasks are resolved. We considered such factors in creating the model depicted in Fig. 1.

Note that the depiction provided in Fig. 1 represents an initial framework; we intend to examine the way teams' Behavioral Concern affects the speed with which various tasks are performed and resolved in a subsequent study. We further operationalize the different components of Fig. 1 in the following section (Section 4).

## 4. RESEARCH METHOD

In order to conduct our study, we examined development artifacts from a specific release (1.0.1) of Jazz (based on the IBM$^R$ Rational$^R$ Team Concert$^{TM}$ (RTC)[1]), a fully functional environment for developing software and for managing the entire software development process [87]. The environment comprises functionality covering work planning and feature traceability, software compositions and builds at various levels of granularity, source code analysis, issue and bug tracking, and version control of artifacts, all in one system [88]. All changes to source code in the Jazz environment are managed solely via the previous creation of work items (WIs), such as a defect, a task, or an enhancement request. Defects in Jazz are actions related to bug fixing, whereas design document development, documentation, or support for the RTC online community are labeled as tasks (although we refer to them here as "support tasks" in order to differentiate from our general use of the term "task" in the paper). Enhancements relate to the provision of new functionality or the extension of existing system features. Team member communication and interaction in relation to WIs are captured by Jazz's comment or message functionality. During development at IBM, project communication, the key content explored in this study, was captured through the enforced use of Jazz itself [37].

The Jazz repository therefore comprised a substantial volume of data collected from distributed software development and management activities across the USA, Canada, and Europe. In Jazz, each team has multiple individual roles with a project leader responsible for the management and coordination of the activities undertaken by the team [89]. All Jazz teams use the Eclipse-way methodology for guiding the software development process [87]. This methodology outlines iteration cycles that are 6–8 weeks in duration, comprising planning, development and stabilizing phases, and generally conforming to agile principles. Builds are executed after project iterations. All information related to the software process is stored in a server repository, which is accessible through a web-based or Eclipse-based (RTC) client interface [36]. This consolidated data storage along with the rigorous project management and control exercised during the project mean that we can place substantial confidence in the reliability and completeness of the data in Jazz.

We applied multiple analysis techniques to the Jazz data in order to answer the RQs above. This multitechnique approach is purported to be useful when exploring phenomena that are not fully understood [90,91], in alignment with the issues under consideration in this work. Standard statistical methods are first used to uncover observable features "at the surface" of the Jazz data, and these outcomes are then triangulated through the use of CA techniques, providing deeper insights into initial patterns detected. We provide details of our data extraction process and metric definitions in the following two Sections 4.1 and 4.2. We then describe our coding process in Section 4.3.

### 4.1. Data extraction

We now briefly address the aspects of data mining that supported the activities involved in this project in terms of extracting, preparing, and exploring the data under observation [92]. Data cleaning, integration, and transformation techniques were used to maximize the representativeness of the data under consideration and to help with the assurance of data quality, while exploratory data analysis (EDA) techniques were used to investigate data properties and to facilitate anomaly detection [93]. Through these latter activities, we were able to identify all records with inconsistent formats and data types, for example, an integer column with an empty cell. We wrote scripts to search for these inconsistent records and tagged those for deletion. This exercise allowed us to identify and delete 122 records that were of inconsistent format. We also wrote and ran scripts that removed all HTML tags and foreign characters from artifacts (as these would have confounded our analysis).

We leveraged the IBM Rational Jazz Client API to extract team information and development and communication artifacts from the Jazz repository. In total, we extracted 30,646 resolved WIs (labeled as one of the three types described above) developed by a total of 474 contributors working on these features between June 2005 and June 2008. These contributors belonged to five different roles: Team leads (or component leads), responsible for planning and executing the architectural integration of components; Admins, responsible for the configuration and integration of artifacts; Project managers (PMC), responsible for project governance; those occupying the Programmer (contributor) role, who contribute code to features; and finally, those who occupied more than one of these roles, labeled Multiple. The WIs were divided among 149 software components (each with functional teams), and 117,101 messages (comments) were exchanged in relation to these WIs. Some teams worked on as few as one WI, while the maximum number of WIs assigned to one team was 4851. The WIs were developed across 30 iterations.

---

[1] IBM, the IBM logo, ibm.com, and Rational are trademarks or registered trademarks of International Business Machines Corporation in the United States, other countries, or both.

As noted above, Jazz project teams were used across locations in North America and Europe; however, we did not consider the specific team location as a variable in this work. We are aware that cultural differences and distance (geographical and temporal) may directly affect software development teams' performance [94], and such conditions might also have an effect on team members' behaviors, which in turn may lead to performance issues [95]. However, previous research examining the effects of cultural differences in global software teams has found few cultural gaps and behavioral differences among software practitioners from, and operating in, Western cultures (the setting for the teams studied in this work), with the largest negative effects observed for teams that spanned Asian and Western cultures [94]. Accordingly, we instead focus on a number of other factors (captured as variables) as derived from the literature around task differences, organizational behavior, and group work in general [37,81–83,89,96,97]; these and other study measures are de- scribed below.

### 4.2. Description of measures

In order to answer the RQs (RQ1, RQ2, and RQ3) introduced in Section 2, we computed a number of metrics and applied statistical analysis methods to the extracted Jazz data. The task, categorized in the repository (and further explained below), was used as our unit of analysis in this work. The dependent variables are team members' participation, team members' engagement, and speed of task resolution. Team members' participation and engagement are considered under the Interaction dimension in Fig. 1 given their relationship. We have initially examined these issues as an intermediate step (team members' participation and engagement) in the first two RQs (RQ1 and RQ2), before answering the last RQ (RQ3). A number of other factors (or predictor variables) are also included in the analysis in order to examine their possible mediation effect on speed of task resolution. We now examine each of these variables and how they were operationalized in turn.

#### 4.2.1. Measuring task type

As noted above, the software tasks of interest here are each categorized as a defect, an enhancement or a support task in the Jazz repository. It was easy to use the in-built task classification scheme from the repository to group and examine software practitioners' undertakings. Although others have previously labeled software features in OSS repositories as either maintenance or nonmaintenance tasks [98], here we examine software tasks as they were actually categorized by those responsible for their delivery – the developers.

#### 4.2.2. Measuring engagement and participation

Those studying software teams have previously operationalized engagement and participation in terms of the number of messages communicated by members [30,53,89,99–101], generally considered under interaction-related factors (refer to Fig. 1). In this work, we also assessed team members' engagement using this metric. In addition, we verified our approach by examining the size (i.e., the length) of the messages exchanged along with the number and roles of the developers involved (discussed further below) during our examination of variance in speed of task resolution. Our basis for taking these steps is that frequency and length of messages may indicate the relative need for idea or information generation, and can generally reflect a team's engagement or information load during team communication [74]; and, participation from a larger cohort of practitioners may indicate the need for knowledge diversity. Diversity here refers to distinctiveness in insights as it relates to the subject under consideration [102]. In fact, use of this measure should help us to understand the uniqueness of the individual tasks in terms of the need for diversity, and hence we studied participation by computing the number of unique developers involved in each software feature (or WI). In addition, when studying participation, we also investigated the contextual situation (i.e., priority) under which tasks were being undertaken (discussed further under (4) Mediators).

#### 4.2.3. Measuring speed of task resolution

Various approaches have been used for many years to measure team- and individual-level performance when undertaking software development tasks. Productivity-related measures such as lines of code per unit of effort [103] and the number of task changes completed [104] are among those used previously to measure performance. Along with others, Cataldo and Herbsleb [104] argued that measures based on lines of code may not be reliable in instances where there is variability in developers' coding styles (e.g., some developers are more verbose than others). In addition, although task changes may be useful for studying performance and the speed at which a developer works [105], this metric is not suitable for studying speed of task resolution. Therefore, given our use of task as the unit of analysis in this work, speed of resolution would be most appropriately measured at the task level [66] (for all tasks that have been resolved). We therefore computed the speed of task resolution of each task by calculating the number of days it took for the task to be resolved. A task was considered to be resolved if its status was set to resolved, closed or verified, and the corresponding date added. This approach has been used elsewhere to measure the speed of task resolution [89,106]. We also considered the associated influence of a number of mediator (predictor) variables during our examination of speed of task resolution, as introduced below.

#### 4.2.4. Mediators

A team's performance is linked to the performance of its individual members. Members' performance in turn is linked to their properties (e.g., traits, characteristics, beliefs, and habits) [26,27]. Thus, enquiries considering team performance are often encouraged to consider group structure. More importantly, given that group work is performed in specific contexts, and focused on particular tasks, this variable (the task) is also said to be a key determinant of the interaction behaviors that are exhibited by teams [25,27,45,48,52] – a position we introduced in Sections 2.1 and 3. In addition, it has been noted that group properties as granular as member age, gender, disposition, beliefs, moods, state of mind and motives, as well as aspects of the physical environment (e.g., noise, heating, and lighting), may also affect team performance [26,27]. In fact, this list represents only a modest subset of potentially influential variables, which may be effectively infinite in number. However, there is limited theoretical basis for the consideration of the majority of these variables, and most are also not easily measured. Of those that are reliably measurable, properties of group structure, situation-related, and interaction-related factors are considered to be important mediators when assessing the outcomes of collective action [25,27,45,48,52,89]. In the same vein, the task under consideration will itself influence group structure [84,85], and the situation under which they are scheduled for resolution [86] (refer to Section 3 for further details). Thus, we organize our mediator (predictor) variables along three dimensions when studying speed of task resolution and the way our variables interacted.

*4.2.4.1. Group structure properties.* Jazz developers were assigned to one of the five roles noted above (with those occupying multiple roles assigned as Multiple in this study). Given that such roles were assigned by upper-level management and that specific intrinsic responsibilities may be assigned to these roles [89], coupled with previous evidence that has established that members' status influences task outcomes [81], we considered the distribution of roles as a potential mediator of speed of task resolution. We validated the mediation effect of group structure on speed of task resolution, and given previous evidence that the task under consideration may influence the group structure that is selected [84,85], we checked for this effect. We therefore computed separately the number of developers, the number of unique roles, and the number of individual roles, to measure group structure.

*4.2.4.2. Situation-related properties.* Tasks with higher priority will generally be delivered sooner than those considered to be less critical [37]. Similarly, tasks developed in certain phases may be delivered with greater urgency than those that are less immediately needed (e.g., those features that are developed closer to a delivery date are likely to be worked on with greater urgency than those developed at the start of a project, or in earlier iterations). We therefore considered the priority of the task and the iteration in which the task was created as situation-related mediators during our assessment of speed of task resolution. We also formally validated earlier assessments that the task may influence the situation and configuration under which teams assemble to resolve them [86], in mediating speed of task resolution.

*4.2.4.3. Interaction-related properties.* In line with previous work [58], we considered a number of communication-related mediation factors associated with team engagement and participation, and particularly those related to communication structures that may influence speed of task resolution (considered under the overarching Interaction mediator – refer to Fig. 1). For instance, although information diversity has been shown to benefit task innovativeness [82,83], and also enhance competitiveness [107], the need to manage a large volume of information can also result in information overload and task delays [96,97]. Under both circumstances, speed of task resolution may be affected. Building on our results derived from answering RQ1 and RQ2 (refer to Sections 5.1 and 5.2), we therefore considered the number of comments and the volume of words communicated in messages around software tasks as potentially influencing speed of task resolution. These metrics were accommodated under the interaction-related factors.

### 4.3. Quantitative CA approach

In order to triangulate the findings drawn from our statistical analysis, we studied in depth 1261 practitioners' messages that were contributed in relation to 250 randomly selected software tasks, using a directed CA approach. We used a hybrid classification scheme adapted from previous works that had examined the details of teams' interactions. The classification schemes of Henri [108] and Zhu [109] are particularly applicable to the work undertaken in this research because of their treatment of the aforementioned subject – the study of which should reveal the nature of practitioners' engagements around software tasks, and the reasons for their interactions. For instance, we should be able to determine if coding a new feature or expending effort on feature enhancements necessitates higher levels of intellectual and cognitive processes than software design tasks.

Use of a directed CA approach is appropriate when there is scope to extend or complement existing theories around a phenomenon [110], and hence suited our further explorations of Jazz practitioners' interactions related to different forms of task. The directed content analyst approaches the data analysis process using existing theories to identify key concepts and definitions as initial coding categories. In our case, we used theories examining knowledge-sharing behaviors expressed during textual interaction [108,109] to support our initial categories (resulting in scales 1–8 in Table 1). Henri [108] and Zhu [109] used Bretz's [111] three-stage theory of interactivity and the group interaction theory of Hatano and Inagaki [112] and Graesser and Person [113], respectively to study teams' interactions. Henri's [108] coding instrument was created to observe five dimensions of interactivity: participative, social, interactive, cognitive, and metacognitive communication, while Zhu's [109] social interaction protocol focused to classify vertical or horizontal interaction. Vertical interaction is characterized by communication where group members seek answers or solutions to problems from capable members, while horizontal interaction involves the strong assertion of ideas, answers, information, discussions, comments, reflections, and scaffolding.

Generally as part of the CA process, should existing theories prove insufficient to capture all relevant insights during preliminary CA coding, new categories and subcategories should be created [110]. Accordingly, both authors of this work and two other experienced coders first classified a random sample of 5% of the 1261 Jazz practitioners' comments in a preliminary coding phase to verify the suitability of the initially created protocol comprising scales 1–8 in Table 1. During this exercise, we found that some aspects of Jazz practitioners' interactions were not captured by the first version of our protocol (e.g., Instructions and Gratitude were missing); the Henri [108] and Zhu [109] protocols did not previously cover these dimensions of interaction. However, we believe that these categories are likely to capture important elements of software practitioners' power systems and their communal (positive) interactions. During this pilot coding exercise, we also found that practitioners in Jazz communicated multiple ideas in their messages. Thus, we segmented the communication using the sentence (or utterance) as the unit of analysis. We extended the protocol, resulting in new scales 9–13 in Table 1 (note: these scales emerged in the order in which they appear in the Table), after which the first author and the two experienced coders coded the 1261 messages that were communicated around the 250 randomly selected tasks. Multiple codes were assigned to utterances that demonstrated more than one form of interaction, and all coding differences were discussed and resolved by consensus. We achieved 81% interrater agreement between the three coders as measured using Holsti's coefficient of reliability measurement (C.R.) [114]. This represents excellent agreement between coders and suggests that a consistent and reliable approach was being taken. The results of this analysis are provided in the following section, subsequent to those derived from our statistical analysis of the artifacts.

Table 1. Coding Categories for Measuring Interaction.

| Scale | Category | Characteristics and Examples |
|---|---|---|
| 1 | Type I Question | Ask for information or requesting an answer – "Where should I start looking for the bug?" |
| 2 | Type II Question | Inquire, start a dialogue – "Shall we integrate the new feature into the current iteration, given the approaching deadline?" |
| 3 | Answer | Provide answer for information seeking questions – "The bug was noticed after integrating code change 305, you should start debugging here." |
| 4 | Information sharing | Share information – "Just for your information, we successfully integrated change 305 last evening." |
| 5 | Discussion | Elaborate, exchange, and express ideas or thoughts – "What is most intriguing in re-integrating this feature is how refactoring reveals issues even when no functional changes are made." |
| 6 | Comment | Judgmental – "I disagree that refactoring may be considered the ultimate test of code quality." |
| 7 | Reflection | Evaluation, self-appraisal of experience – "I found solving the problems in change 305 to be exhausting, but I learnt a few techniques that should be useful in the future." |
| 8 | Scaffolding | Provide guidance and suggestions to others – "Let's document the procedures that were involved in solving this problem 305, it may be quite useful." |
| 9 | Instruction/Command | Directive – "Solve task 234 in this iteration, there is quite a bit planned for the next." |
| 10 | Gratitude/Praise | Thankful or offering commendation – "Thanks for your suggestions, your advice actually worked." |
| 11 | Off task | Communication not related to solving the task under consideration – "How was your weekend?" |
| 12 | Apology | Expressing regret or remorse – "Sorry for the oversight and the failure this has caused." |
| 13 | Not Coded | Communication that does not fit codes 1–12. |

## 5. RESULTS

We provide our results in this section. Section 5.1 first details the results from our initial statistical analysis involving the artifacts, separated into three subsections addressing RQ1–RQ3 in turn. The second set of results then follows in Section 5.2, reporting our quantitative CA directed at triangulation and elaboration of the initial findings in Section 5.1.

### 5.1. Initial results

#### 5.1.1. Software tasks' distribution and practitioners' participation (RQ1)

Out of >30,000 software tasks that were extracted from our instance of the Jazz repository by far the largest group have defects, comprising 23,331 tasks (76.1%). A further 12.2% (3748 of the 30,646 tasks) were classified as support tasks, and the remaining 11.6% (3567) were classified as enhancements (providing new functionality or the extension of system features). Although there was overlap in team members addressing each form of task, overall, work on defects required input from the largest cohort of practitioners (411 practitioners or 86.7% of the 474 members), enhancements were worked on by the second highest number of practitioners (226 practitioners or 47.7%), and support tasks involved slightly fewer members (212 practitioners or 44.7%). We were also able to consider the extent to which developers conducted work on each type of task, providing a finer grained view of participation intensity; relevant summary statistics are presented in Table 2. These results show that, on average, more Jazz practitioners worked together to address defects and enhancements than they did to undertake support tasks.

Although the absolute differences in mean values shown in Table 2 are small, we were interested in knowing whether these differences were statistically significant. Given the large sample size, we first used the Kolmogorov–Smirnov test to check the normality of practitioners' participation in the three forms of task (refer to Brooks et al. [115] for discussions on the formal application of normality tests). The results of these tests confirmed that the data distributions for all the three categories of task deviated significantly from a normal distribution ($p < 0.05$). The standardized skewness coefficient (i.e., the skewness value divided by its standard error) and standardized kurtosis coefficient (i.e., the kurtosis value divided by its standard error) were also outside the boundaries of normally distributed data (i.e., -3 to +3) [116]; refer to Table 2 for details. Thus, the nonparametric Kruskal–Wallis test was used to test for differences in practitioners' participation in the three forms of task. This test revealed a statistically significant difference in the extent to which practitioners' participated in the three types of tasks $X^2 = 190.733$, $p < 0.001$, with a mean rank of 15607.01 for defect, 15294.51 for enhancement, and 13586.29 for support task. The effect

Table 2. Descriptive statistics for Jazz practitioners' participation in different forms of software tasks.

| | Mean | Median | Std Dev | Min | Max | SE (Mean) | SK | KS | SE (SK) | SE (KS) |
|---|---|---|---|---|---|---|---|---|---|---|
| Defect | 2.0 | 2 | 1.2 | 1 | 19 | 0.01 | 2.0 | 8.8 | 0.02 | 0.03 |
| Enhancement | 2.0 | 2 | 1.2 | 1 | 10 | 0.02 | 1.9 | 5.6 | 0.04 | 0.08 |
| Support Task | 1.8 | 1 | 1.2 | 1 | 13 | 0.02 | 2.5 | 10.5 | 0.04 | 0.08 |

Note: Std Dev = Standard Deviation, SE = Standard Error, SK = Skewness, KS = Kurtosis.

size associated with this difference, as measured by Cramer's V, was small, 0.10. A series of three Mann–Whitney pairwise follow-up tests at the Bonferroni-adjusted level [117] of 0.016 (i.e., 0.05 divided by 3 analyses) indicated that significantly more Jazz practitioners worked on defects than support tasks ($p < 0.016$); and a similar result is obtained for the difference between enhancement features and support tasks ($p < 0.016$). By contrast, practitioners' participation in defects and enhancement tasks did not reveal a significant difference ($p > 0.016$).

Although over minimal magnitudes, these outcomes were duplicated when we examined how those occupying particular roles participated in the three forms of tasks. Our checks to see whether task priority affected the pattern of results above (i.e., if there was a covariance effect) also confirmed that this was not the case. In fact, on average, support tasks had the highest priority (mean = 2.6, median = 3.0, std dev = 1.2), followed by enhancement features (mean = 2.2, median = 2.0, std dev = 1.2), then defects (mean = 2.1, median = 1.0, std dev = 1.2) – where "4" is the highest possible priority and "1" is the lowest.

**5.1.2. Covariance in team members' engagement (RQ2)**
Of the 117,101 messages that were exchanged around the 30,646 tasks, 88,874 messages (or 75.9%) were exchanged by practitioners working on defects, 14,512 messages (12.4%) were exchanged by practitioners working on enhancement features, and 13,715 messages (11.7%) were exchanged by practitioners working on support tasks. This spread reflects a generally similar (and hence not unexpected) pattern to the distribution of the actual software tasks noted above. We take a finer grained view of the descriptive statistics for the numbers of messages that were communicated when members were resolving software tasks in Table 3. Here it is shown that the pattern of results is different to the overall aggregated results. For instance, descriptive statistics in Table 3 show that practitioners engaged most intensively when they were working on enhancement features, exchanging on average 4.1 messages for each task. The second highest number of messages was typically exchanged when practitioners were working on defects (3.8 messages on average), while the fewest messages were generally exchanged when practitioners were addressing support tasks (3.7 messages on average). As mentioned above, we conducted formal statistical testing to determine whether the differences noted in Table 3 were statistically significant. We first examined the distributions for the numbers of messages exchanged by practitioners when they were working on the three forms of task. Kolmogorov–Smirnov tests confirmed that the data distribution for all the three categories of task significantly deviated from a normal distribution ($p < 0.05$ for all three tests). Therefore, a Kruskal–Wallis test was used to check for differences in practitioners' engagement in the three forms of task. We normalized our engagement metric (number of messages communicated) based on the number of developers, on the backdrop that more developers associated with specific software tasks will generate more messages, which would affect the reliability of our results. This normalized result revealed a statistically significant difference in the way practitioners' engaged when solving the three forms of task $X^2 = 19.833$, $p < 0.001$, with a mean rank of 15895.60 for enhancement, 15384.71 for support task, and 15226.20 for defect. The effect size associated with this difference, as measured by Cramer's V, was small, 0.02. A series of three Mann–Whitney pairwise follow-up tests at the Bonferroni-adjusted level of 0.016 indicated that Jazz practitioners engaged significantly more around enhancement features than defects ($p < 0.016$), and on enhancement features than support tasks ($p < 0.016$). However, there was no difference in practitioners' engagement when they were working on support tasks and defects ($p = 0.295$).

**5.1.3. Variance in speed of task resolution (RQ3)**
We present our results regarding differences in speed of task resolution for the three forms of task in this section. Similar to the procedures outlined above, we first examined the descriptive statistics for the time taken to undertake the three forms of task. These results show that enhancement tasks took the longest time to complete (in days, mean = 67.0, median = 19.0, std dev = 109.7), support tasks took the second highest number of days (mean = 42.1, median = 11.0, std dev = 78.7), while defect tasks were completed most quickly (mean = 43.5, median = 5.0, std dev = 98.1). The data distributions for all the three forms of task violated the assumption of normality. We therefore conducted a Kruskal–Wallis test to check for differences in speed of task resolution for the different forms of task. This result revealed a statistically significant difference in speed of task resolution $X^2 = 677.853$, $p < 0.001$, with a mean rank of 18360.18 for enhancement, 16707.37 for support task, and 14636.92 for defect. The effect size associated with this difference, as measured by Cramer's V, was small, 0.11, but higher than those noted above (for RQ1 and RQ2). As executed above in Sections 5.1.1 and 5.1.2, a series of three Mann–Whitney pairwise follow-up tests at the Bonferroni-adjusted level of 0.016 confirmed that practitioners spent significantly more time addressing enhancements than they did for defects ($p < 0.016$) and support tasks ($p < 0.016$), respectively. Similarly, there was a statistically significant difference between the number of days taken to address support tasks and defects ($p < 0.016$).

We next tested the effect of the mediator variables that were included in Fig. 1 (i.e., those related to Group Structure, Situation, and Interaction). Baron and Kenny [118] recommended estimating a number of regression equations to test for mediation effect: (1) regressing the independent variable on the mediator variable, (2) regressing the independent variable on the dependent variable, (3) regressing the mediator variable on the dependent variable,

Table 3. Descriptive statistics for Jazz practitioners' engagement in different forms of software tasks.

|  | Mean | Median | Std Dev | Min | Max | SE (Mean) | SK | KS | SE (SK) | SE (KS) |
|---|---|---|---|---|---|---|---|---|---|---|
| Defect | 3.8 | 2.0 | 4.5 | 1 | 266 | 0.07 | 11.5 | 50.0 | 0.02 | 0.03 |
| Enhancement | 4.1 | 3.0 | 4.8 | 1 | 57 | 0.08 | 3.9 | 23.2 | 0.04 | 0.08 |
| Support Task | 3.7 | 2.0 | 4.5 | 1 | 74 | 0.03 | 4.4 | 33.5 | 0.04 | 0.08 |

Note: Std Dev = Standard Deviation, SE = Standard Error, SK = Skewness, KS = Kurtosis.

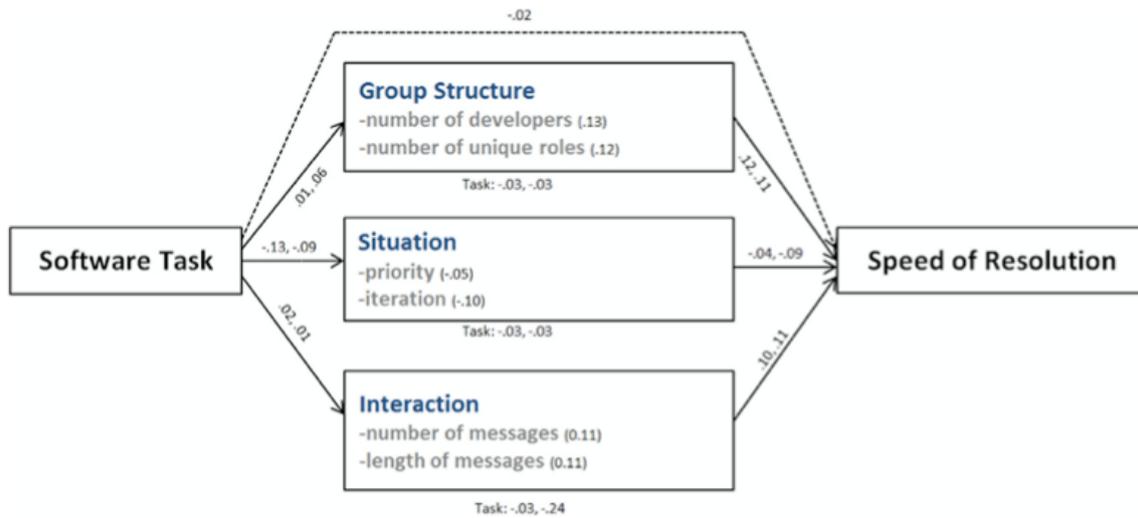

Fig. 2. Model of the relationships among software tasks, mediators, and speed of resolution (results are all significant to at least a p < 0.05 level).

and (4) regressing the independent and mediator variables on the dependent variable (multiple regression). They noted that in order to establish mediation: (1) the independent variable must affect the mediator in the first test, (2) the independent variable must affect the dependent variable in the second test, and (3) the mediator variable must affect the dependent variable in the third test. In addition, these authors advised that the independent variable should have less effect on the dependent variable in the fourth equation than the second equation, where perfect mediation is established if the independent variable has no effect when the mediator is controlled [118].

We have followed Baron and Kenny's [118] recommendations in testing the preliminary model presented in Fig. 1, and Fig. 2 provides our updated model with coefficients included. These results are all significant at the p < 0.05 level. In addition, in Fig. 2 beta coefficients from our multiple regression analysis for the "Software Task" variable are included below the boxes in the model, with multiple regression outcomes recorded in the order in which the variables appear in the box. Beta coefficients for the variables included under the main mediators (e.g., Group Structure) are listed beside the actual variables that were considered (e.g., number of developers) for the multiple regression outcomes. Beta coefficient are included along the paths in the order that variables appear in the boxes for the simple regression outcomes (e.g., number of developers = 0.01 and number of unique developers = 0.06 for the path between "Software Task" and "Group Structure").

In Fig. 2, it is shown that for the Group Structure mediators, the first three conditions of Baron and Kenny [118] were established, and the independent variable (Task type) had a marginally stronger effect in the final multiple regression equation (beta = 0.03) than in the second equation (beta = 0.02) for each of the variables (Number of developers and Number of unique roles). A similar pattern was also observed for Number of individual roles, not included in the figure. In addition, we observed slight increases in the mediator effect in the last multiple regression equations for all of the Group Structure mediators in Fig. 2. These outcomes support partial mediation – albeit with low magnitudes.

Fig. 2 depicted a similar pattern of outcomes for the Situation- related mediators (Iteration and Priority). Here it is seen that the independent variable (task type) had a slightly larger effect on the dependent variable (speed of resolution) in the last multiple regression equation (beta = 0.03 when Iteration is controlled and beta = 0.03 when Priority is controlled, respectively, compared with beta = 0.02 in the second equation). Furthermore, our outcomes for the Interaction-related mediators duplicated those for Group Structure, where all conditions suggested by Baron and Kenny were considered and the independent variable (Task type) had a larger effect in the last multiple regression equation (beta = 0.03 and 0.24) than in the second equation (beta = 0.02) for each of the variables (Number of messages and Length of messages). Our results here also support partial mediation. Given these outcomes, statistical analysis was used to build a regression model to

Table 4. Correlations among the dependent and predictor variables.

| Variable | 1 | 2 | 3 | 4 | 5 | 6 | 7 | 8 | 9 | 10 | 11 | 12 | 13 |
|---|---|---|---|---|---|---|---|---|---|---|---|---|---|
| 1 Time taken | 1.0 | 0.12 | 0.11 | 0.06 | 0.01 | 0.15 | −0.02 | 0.09 | −0.09 | −0.04 | 0.10 | 0.11 | −0.02 |
| 2 Number of developers | | 1.0 | 0.79* | 0.45 | 0.09 | 0.34 | 0.47 | 0.24 | 0.05 | 0.08 | 0.83* | 0.69* | 0.01 |
| 3 Number of roles | | | 1.0 | 0.38 | 0.18 | 0.37 | 0.34 | 0.32 | 0.02 | 0.06 | 0.65* | 0.53* | 0.06 |
| 4 Team lead | | | | 1.0 | −0.03 | −0.15 | −0.26 | 0.02 | 0.00 | 0.11 | 0.37 | 0.32 | −0.05 |
| 5 Admin | | | | | 1.0 | −0.04 | −0.11 | 0.01 | 0.00 | −0.06 | 0.05 | 0.01 | 0.02 |
| 6 Project manager | | | | | | 1.0 | −0.01 | 0.04 | −0.06 | −0.05 | 0.29 | 0.23 | 0.07 |
| 7 Programmer | | | | | | | 1.0 | −0.07 | 0.13 | 0.07 | 0.40 | 0.32 | 0.06 |
| 8 Multiple | | | | | | | | 1.0 | −0.06 | −0.01 | 0.20 | 0.19 | 0.04 |
| 9 Iteration | | | | | | | | | 1.0 | 0.36 | 0.09 | 0.06 | −0.09 |
| 10 Priority | | | | | | | | | | 1.0 | 0.11 | 0.11 | −0.13 |
| 11 Number of comments | | | | | | | | | | | 1.0 | 0.83* | 0.02 |
| 12 Message length | | | | | | | | | | | | 1.0 | 0.01 |
| 13 Task type | | | | | | | | | | | | | 1.0 |

Note: *strong statistically significant + correlation (p < 0.001); dummy dichotomous variables replaced the continuous "Task type" in our model given that there were three categories of tasks – defect, enhancement, and support task (above, support tasks were coded as 1, enhancement tasks as 2, and defect tasks as 3).

Table 5. Results from regression analysis.

| | Unstandardized Coefficients | | Standardized Coefficients | t | Sig. (p) |
|---|---|---|---|---|---|
| | B | Std. Error | Beta | | |
| (Constant) | 38.48 | 2.10 | | 18.37 | 0.000 |
| **Project Manager** | 50.42 | 1.74 | 0.16 | 28.93 | 0.000 |
| **Multiple** | 40.84 | 2.67 | 0.09 | 15.31 | 0.000 |
| **Team Lead** | 21.23 | 1.31 | 0.10 | 16.19 | 0.000 |
| **Iteration** | −7.26 | 0.52 | −0.08 | −13.90 | 0.000 |
| Enhance | 19.74 | 2.23 | 0.07 | 8.83 | 0.000 |
| Admin | 16.07 | 3.52 | 0.03 | 4.57 | 0.000 |
| **Programmer** | 5.95 | 1.35 | 0.03 | 4.43 | 0.000 |
| Defect | −5.61 | 1.70 | −0.02 | −3.30 | 0.001 |
| **Priority** | −2.76 | 0.97 | −0.02 | −2.86 | 0.004 |

Note: **bold** variables were log transformed.

examine whether/how the variables affected speed of task resolution, and our overall pattern of results. More generally, our model also enables us to understand the way specific factors (including our mediators) might interplay when practitioners were working to undertake different software tasks. Given that the distributions for the measures were all skewed, we first performed a natural log transformation on the variables. We then performed Pearson product-moment correlation tests in order to examine how the variables were related. This exercise was also used to support the selection of relevant variables for our model, in order to avoid multicollinearity. The correlation matrix is provided in Table 4 (noting that "Task type" is presented as a continuous variable, although its replacement as three dummy variables did not affect our model or the overall outcomes). Understandably, our results confirmed that the numbers of roles, comments, and message length increased when there was an increase in practitioner numbers working around tasks. In addition, practitioners communicated longer messages when there were more exchanges. Of note in the correlation matrix, however, none of the variables had a major effect on speed of task resolution (refer to Table 4) – there were small correlations between time taken and number of developers, number of roles, presence of project managers, number of comments, and message length. This finding runs counter to those that were previously uncovered by researchers observing other teams [58]. In terms of our regression model, although a significant model emerged ($F_{9,30636} = 197.892$, $p < 0.001$) our adjusted R-squared value revealed that our model only accounted for 6% ($R^2 = 0.055$) of the variance in speed of task resolution. The beta coefficients for the significant variables ($p < 0.01$) include: Project Manager = 0.16, Team lead = 0.10, Multiple = 0.09, Iteration = 0.08, Task type = 0.05 (Enhancement = 0.07 and Defect=0.02), Admin=0.03, Programmer=0.03, and Priority = 0.02. We provide further regression results in Table 5. Our outcomes here show that, although there were mild variances in speed of task resolution when practitioners were addressing the three forms of task given the effects of various factors (with Group Structure having the most significant effect), our variables (including the task itself) were not major predictors of these differences.

Given the inherent limitations associated with the quantitative techniques used here [119–121], and the rather inconclusive results just reported, we therefore contextually analyzed a sample of the practitioners' engagements with a view to triangulating our findings and providing more circumstantial evidence around the practitioners' behavioral processes when addressing different forms of task. These results are provided in the following section.

### 5.2. Quantitative CA results

As noted in Section 3, we randomly selected 250 tasks in order to take an in-depth look at developers' participation and engagement around these tasks using a bottom-up analysis approach. We selected 150 defects, 50 enhancements, and 50 support tasks, in line with the relative prevalence of tasks in the overall distribution. In total, 1261 messages were communicated around these three forms of task: 738 messages relating to defects, 294 messages for the support tasks, and 229 messages around enhancements. From these 1261 messages, 4136 codes were recorded. Table 6 provides summary counts of these codes. In Table 6, it is of note that Information Sharing, Discussion, Scaffolding, and Comments dominated Jazz practitioners' exchanges across all the three forms of task – 76.1% of all codes were recorded to these four categories. In comparison, there were fewer Answers, Type II Questions, Instructions, and Reflections (16.8% of all codes were recorded to these categories). In addition, a much lower proportion of codes are observed for Off task, Type I Questions, and Gratitude utterances in Table 6 (6.7% of all codes were recorded to these categories). Finally, just 13 and 6 utterances were coded to Apology and Not Coded, respectively (0.5% of all codes).

Pearson Chi-square tests were conducted to ascertain whether the differences observed in Table 6 are statistically significant. This statistical procedure is considered to be appropriate when the distributions comprise frequency data, as is the case for the codes that were obtained for our analysis of the 1261 messages through the directed CA process [122]. Furthermore, with the exception of the Apology and Not Coded categories, all the data samples comprised a sample size that was substantially >10 (the assumption for using a Chi-square test) [122].

Table 6. Interaction categories (utterances) and number of occurrences for defects, enhancements, and support tasks.

| Category | Defect | Enhancement | Support Task | ∑ |
|---|---|---|---|---|
| Type I Quest. | 63 | 15 | 14 | 92 |
| Type II Quest. | 108 | 37 | 46 | 191 |
| Answer | 119 | 38 | 47 | 204 |
| Information Sharing | 1133 | 383 | 391 | 1907 |
| Discussion | 227 | 113 | 139 | 479 |
| Comment | 182 | 70 | 65 | 317 |
| Reflection | 82 | 22 | 22 | 126 |
| Scaffolding | 266 | 93 | 85 | 444 |
| Instruction/Command | 85 | 28 | 59 | 172 |
| Gratitude/Praise | 49 | 21 | 18 | 88 |
| Apology | 9 | 3 | 1 | 13 |
| Off task | 61 | 15 | 21 | 97 |
| Not Coded | 2 | 0 | 4 | 6 |
| ∑ | 2386 | 838 | 912 | 4136 |

The results of the Chi-square test confirm that there were significant differences in the types of utterances exchanged by Jazz practitioners, and particularly for the higher levels of Information Sharing, Discussion, Scaffolding, and Comments that were communicated ($X^2 = 12408.0$, df = 9, $p < 0.001$). Given our objective to understand further how Jazz practitioners participated and engaged when resolving various forms of software task (and to triangulate the manifest findings above), we examine these codes across the three forms of task in Fig. 3.

In light of the differences in messages that were communicated, and the subsequent codes that were derived for the different forms of task in Table 6, we normalize the codes across tasks by using percentages in Fig. 3 (codes for the Apology and Not Coded categories are omitted due to the small number assigned to these categories). Fig. 3 shows that practitioners working on defects typically asked slightly more questions than those working on enhancements and support tasks (2.6% compared with 1.7% and 1.5%, respectively). In general, however, Questions and Answers were distributed relatively evenly across all three forms of task. However, there were also higher proportions of Information Sharing on defects and enhancements than support tasks (47.5% and 45.7%, respectively, compared with 42.9% for practitioners working on support tasks). By contrast, Fig. 3 demonstrates that there was proportionally more Discussion around support tasks (15.2%), while those working on enhancements communicated the second-highest level of this form of language (13.5%), and defects saw the lowest extent of Discussion (9.5%). Comments were most prevalent around enhancements and defects (8.4% and 7.6%, respectively), with those working on support tasks using a lower level of this form of language (7.1%). Use of Reflections was even more, although members working on defects tended to reflect the most (3.4% compared with 2.6% and 2.4% for those undertaking enhancements and support tasks, respectively). Fig. 3 further illustrates that while Scaffolding was even for practitioners working on defects (11.2%) and enhancements (11.1%), this type of utterance was less evident for those working around support tasks (9.3%). By contrast, practitioners were instructed almost twice as often when they were working around support tasks (6.5%) compared with defects (3.6%) and enhancements (3.6%). Gratitude and Off-Task interactions were much more even across the three forms of task.

Following a similar approach to that used above, we examine the percentage interactions for significant differences across the three forms of tasks using a Chi-square test. The results of this test show that the differences noted in Fig. 3 above were indeed statistically significant, supporting the quantitative results that practitioners' participation and their engagements covary given the type of task they were performing ($X^2 = 62.557$, df = 24, $p < 0.001$). These findings may have implications for software task assignment and team composition in general. We examine such analyses in the following section.

## 6. DISCUSSION

*RQ1. (a) How are software tasks distributed in a large software project, and (b) how do team members participate in the different ecosystems of tasks?* Mostly, during the 3-year period considered in our analysis, Jazz developers were mostly occupied with fixing bugs. Team members were also required to address more software design, documentation, or support tasks rather than developing new features and software application extensions. In terms of members' roles and their overall participation, a wider spread of members (and roles) assembled to fix defects and to work on delivering software enhancements and new features than they did for design, documentation, and support tasks. This is despite the fact that these latter tasks had typically been assigned much higher priorities. Although it may be typical for a smaller spread of members to be involved in software requirements gathering and design tasks, wider team involvement in these activities may benefit project performance, and particularly given the volume of defect work relative to other tasks noted in this study. In fact, difficulty in treating defects is held to reduce when there is good understanding of feature requirements and design [123], which may be evidenced by practitioners' participation and their involvement in a distributed development context. Perhaps a greater spread of developers' involvement in earlier project activities would therefore help developers to reduce the effort required to resolve defects. Those defects could be of many types (e.g., an error in the program code, software not functioning as was originally intended in the proposed requirements, the user guides not matching the business process, and so on); of particular interest, however, are those defects that result through the development of requirements that are not entirely understood. Although we were not able to fully differentiate these forms of software issues from other forms of defects, this evidence for the higher prevalence of defects and the lower level of developer participation in design, documentation, and support tasks suggest that wider participation in the aforementioned activities, and particularly activities related to requirements gathering, may benefit such teams and reduce the incidence of potential defects.

Of course, the higher levels of member participation in addressing defects, new features, and software extensions may be a deliberate strategy by Jazz project management. Alternatively, this may be because of the evidence of self-organization to maintain team performance [124,125]. The former assessment is indeed seemingly plausible given the evidence that design, documentation, and support tasks benefitted from higher prioritization, which would typically be set by top (senior) management. Higher priority tasks are generally given more attention, and particularly in terms of scheduling and member participation. However, our results do not support this generally established pattern. Rather, our evidence tends to support the latter view that these teams could have been effectively self-organizing given both schedule pressures (and hence the need to dedicate more resource to fixing defects, new features, and extensions) and their perception of support and documentation tasks being less relevant (in contravention of management priorities). This behavior is not uncommon among developers who may circumvent

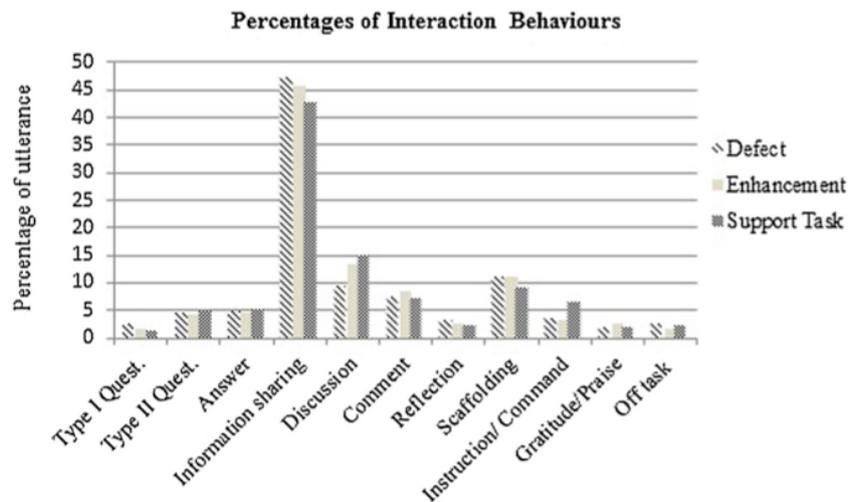

Fig. 3. Percentages of practitioners' interactions (utterances) related to defects, enhancements, and support tasks.

senior level instructions at times to self-organize and maintain speed in task resolution [126].

*RQ2. Do the level and type of team members' engagement covary with the nature of the task they are performing?* The Jazz developers in our study typically expended relatively more effort engaging with one another regarding new features and software extensions. On defects, design, documentation, and support tasks, these practitioners had the fewest exchanges. Notwithstanding that some aspects of practitioners' communication with clients and users may not be captured in the artifacts we explored, to us this was an initially unexpected result, given that crafting, delineating, and clarifying new requirements (labeled under the support task category) are generally perceived to necessitate high levels of verification and communication to ensure adequate decomposition of the application domain, even among developers. Of course, documentation related to software architecture and end-user support (and those similar in nature, e.g., "user instructions" and "how to" guides) may not necessarily demand significant levels of communication. In addition, those undertaking the latter forms of task may be relatively familiar with the perceived workings of software features, or may be given specific instructions for the crafting of end-user support documentation based on already developed software, and hence, may engage less. By contrast, higher levels of communication around new feature development tasks is expected in an agile development context (as promoted in the Eclipse-way approach used by these Jazz teams [87]), where requirements gathering and clarification is an ongoing process, and communication and collaboration are valued over documentation. Thus, these findings for a higher level of engagement around new features converge somewhat with evidence established in the literature.

Perhaps our results in this work may be linked to the specific agile approach implemented by these teams for developing software features. Those operating in the agile paradigm stress the need for iterative development with multiple feedback loops, where system requirements evolve [127,128]. We observe the highest degree of engagement during feature development. Thus, the higher level of communication during feature developments seems to demonstrate some level of convergence between what is stressed in theory [127] and the evidence in the artifacts of these Jazz practitioners. A strategy to develop features rapidly and then spend (possibly more) time perfecting them may find the right balance between time taken to deliver and customer satisfaction. However, too high prevalence of bugs may not be pleasing to project management if considered in isolation. This occurrence may in fact be perceived as poor or inadequate performance.

Given previous evidence, which shows that software developers often consider defects (requiring bug fixes) as team obstacles [129], we expected that there would be a lesser degree of engagement on such tasks. We anticipated that, given the likely negative attitudes to such tasks, developers involved in this form of activity would demonstrate reduced communication patterns when compared with other forms of task, with members just wanting to get bug-related issues "out of the way." In addition, given that defects are typically discovered after features are coded (during testing), we anticipated that specific developers occupied with these tasks would already have a sound understanding of the necessary workings to address such issues, and so may not necessarily need a large amount of help from others. Furthermore, we suspected that more effort would be expended identifying bugs than in applying the fix itself. Thus, there would be a lesser need for a large amount of engagement when fixing bugs.

Our findings indeed confirm a lower level of engagement around defects, somewhat supporting these propositions. However, overall, 86.7% of all team members were engaged in addressing defects, a much higher proportion than for those involved in providing new features and software extensions, and in design, documentation, and support tasks. As noted above, notwithstanding that this may be a deliberate strategy (i.e., to have many team members involved in bug fixing) used by Jazz teams given short release cycles and the demand to deliver features, these results contribute support to the use of approaches that reduce duplicate bug reports [130] and that lessen such issues arising in the first instance [131]. Continuous integration practices may also support rapid detection of defects, and hence, encourage early fixing of such issues. Reducing duplicate bug reports should decrease the time spent unnecessarily on issue resolution, while test-driven development (TDD), continuous integration practices, and similar early-intervention approaches can improve software quality, also lessening bug incidence [130,131]. In fact, as noted above, expending greater effort on requirements gathering and decomposition and software design, particularly involving a large cohort of developers, may enhance developers' performance and reduce the incidence of defects. Furthermore, time spent on searching for bugs, estimating and planning work effort, and fixing these issues, could instead be directed to enhancing software artifacts and the delivery of new features.

Evidence uncovered from our contextual analysis indeed confirms that time spent answering questions and sharing information during bug fixing could be used otherwise (refer to Section 5.2 for details), providing another lens for interpreting our findings. Although we noted some similarities in the way software practitioners interacted when they were addressing different forms of software task, our results in this study show that the nature of practitioners' engagements generally covaried with task, and particularly for the level of information, discussions, scaffolding, and instruction they communicated. For instance, practitioners working on defects generally shared more information and their engagement comprised more scaffolding, particularly when compared with those working on support tasks. However, this pattern of interaction was reversed for discussions (or communications aimed at generating and exchanging ideas) and instructions, where those working on support tasks communicated more often through these forms of interaction. Likewise, more comments (i.e., communication of a judgmental nature) were exchanged when practitioners were developing new software features or extending software functionality.

Evidence above showing lower levels of engagement around defects is fitting; however, the larger amounts of information and scaffolding that were communicated by those addressing defects are surprising. The more common incidence of these forms of utterance corresponds with the slightly higher extent of questions that were asked by these individuals. Of course, given that there were fewer developers addressing new features and extending already developed software functionalities, it was fitting that when bugs resulted from these features any new members who were responsible for providing appropriate solutions would lack context. Accordingly, these new practitioners would need more information than was typical for members working on the other types of task (e.g., new features). In fact, our contextual results also support our proposition above for the potential negative effect of the lack of widespread participation in design and documentation tasks. Although we are not able to definitively link this lack of participation in design and documentation tasks to the higher prevalence of defects as such, those resolving defects appeared to compensate for the higher number of team members' requests for help by providing more guidance and suggestions through scaffolding. Coordination theories have shown that task programming (i.e., setting up mechanisms to automatically address repetitive tasks through tools, documentation, and specifications – as seen for those engaged in scaffolding utterances) reduces the need for team engagement [94], and this attempt was evident among those addressing defects. Members resolving defects tended to anticipate future enquiries, and thus, made provision for these through task-programming behaviors. Other intuitions not evident in the artifacts studied may also contribute to practitioners' scaffolding behaviors while addressing defects. For instance, with an organization culture to assemble larger teams to address defects, scaffolding may thus be encouraged (or required), forming a team norm.

On the basis of evidence uncovered by those examining group interactions and teams involved in different forms of task [25,27,45,48,52], we anticipated that members who were involved in software design, documentation, or support tasks would benefit from higher levels of instructions and cooperation [47]. Previous work had indeed observed those involved in requirements gathering and software design to be most collaborative [75]. In addition, we anticipated that increased consensus [76] would benefit those operating on documentation and software support and design tasks. Our contextual analysis indeed revealed that such members elaborated, exchanged, and expressed ideas or thoughts more freely. We also found practitioners working around these tasks to be instructed the most. This finding is particularly interesting when considering that these members asked the fewest questions and were least judgmental. Notwithstanding the relatively small sample size that was used for the contextual analysis, we believe that these are interesting and potentially useful results. These findings may have particular implications for practice. For instance, the higher level of instructions may have implications for the availability of leading team members as junior members may grow depending on their instructions, which may delay work in their absence. In addition, team environments in which members freely express themselves may have implications for the availability of such open members, who may not be in abundant supply.

In fact, our findings regarding practitioners' interaction patterns when they were engaging on new software features and feature enhancements show that these members also shared large amounts of information, ideas, and scaffolding, but they were most judgmental. This latter form of team process is said to be good for enhancing innovativeness and critical evaluation among group members [132], processes that are likely to be useful during software coding. In addition, value diversity among teams has also been shown to increase this form of behavior [82]. Considering all of the results together for those working on new software features and feature enhancements, it is shown that these members expressed most intellectual and cognitive processes [45]. These initial insights are useful and show promise for further large-scale explorations.

*RQ3. (a) Are there variances in speed of resolution for different forms of software task, and (b) what factors affect speed of resolution of software tasks?* Our results show that Jazz practitioners typically spent the longest developing new features and providing software enhancements. Design, documentation, and support tasks also took marginally longer to resolve than defects. This finding is particularly fitting given that new features and software extensions generally demand intellectual and cognitive

processes [20], that are likely to mature over time, when compared with defects that relate to features that have been already developed. Thus, practitioners are expected to have some amount of previous knowledge about the relevant software feature when addressing bugs. Evidence that defects were completed more quickly is also understandable for these teams, particularly given that such features dominated developers' activities and involved more team members. Furthermore, defects may be perceived to require less effort, which may encourage practitioners to schedule and fix such tasks earlier. The lack of familiarity for the larger cohort of practitioners working on defects could also have been expected to negatively influence the speed of task resolution (and the time taken to locate bugs), but this was not the case. In fact, we found that, overall, task completion took slightly longer when there were larger cohorts of developers involved, and particularly when the team comprised more project managers. In addition, software tasks in general took longer when there were more messages exchanged. The effects of message overload on team speed of task resolution has been noted previously [96,97], although our results were not highly revealing in this study.

Overall, although previous work, and particularly studies dedicated to understanding software development teams' pro- cesses, has observed that the number of developers, feature size, response time, and interaction have a significant effect on the duration of software feature development [37,58], we did not find these types of factors to have a major effect on Jazz practitioners' speed of task resolution in this work. Instead, we anticipate that behavioral and intrinsic issues may interact with these extrinsic variables becoming more significant predictors of speed of software task resolution. Our contextual analysis has indeed provided initial insights into Jazz practitioners' behavioral processes that somewhat endorse this viewpoint (see differences in the nature and forms of utterance for those addressing the range of tasks, in Fig. 3). Given our findings in this work, we believe that, although software tasks possess various properties that are likely to influence those performing them, and hence, the speed with which tasks are performed, the complexities inherent in human decisions and attitudes may instead have a greater effect on the speed of task resolution. In fact, although most software development tasks do not entirely conform to the distinctions (classifications) in established group task models [25,27,45,48,52], there is convergence on the relevance of human behavior and attitudes in understanding task performance, and preliminary findings considering software teams have indeed observed differences in the behaviors expressed by those working on different forms of features [20].

For instance, the intellectual and creativity tasks of Laughlin [51] and Hackman [25] outline those activities that are of a cognitive nature where there is an obvious right answer, or intelligent tasks that are solved based on social consensus. According to Laughlin [51], apart from tasks with demonstrably correct answers (e.g., logic problems), the solution to all other cognitive tasks is driven by compromise and reasoning among those undertaking them (e.g., tasks involving negotiation). Most software development tasks are creative in nature [73], and fit this latter profile, where the success indicator for delivery of a software task is not static, tending to be shaped by compromise among project stakeholders. For example, a banking client may request software to manage loans from a software supplier, comprising a feature for tax and insurance accounting (assuming that the developers do not decompose such a feature into multiple units of work, or subfeatures, e.g., several user stories). In this scenario, the success indicator for the client (and software supplier) would be the availability of both features in the software. The software project organization – the supplier – may schedule this feature in a specific time frame, which may be eroded with the team only addressing a part of the software functionality, for example, the tax accounting aspect. In this instance, although the success indicator of the feature was initially the availability of the tax and insurance accounting features, the client contact at the software project organization may manage to convince the client to accept the partially completed project in the first instance, while the other aspect related to insurance accounting is developed for a later release. With the client's acceptance, this feature would possibly be signed off by the development team and client as successfully resolved, in a way shifting the success indicator. In fact, previous work has indeed noted this willingness by clients to perform limited software evaluation, where these members were keen to provide the maximum score for satisfaction [133]. Of course, a host of factors, including culture and other behavioral issues (e.g., mood [14,134] and personality [135]) are likely to come into play during such a scenario, where some clients may not be so easily persuaded. Perhaps the software project organization contact may use examples of customers that accepted a similar partial feature as a form of anchoring. They may also provide evidence of previous customers starting to receive immediate returns on their investment, and so on, all toward convincing the client to agree to the incremental release of the feature, and thus, shifting the success indicator.

In fact, the aforementioned example reflects a very simplistic scenario of the effects of social consensus on the speed of task resolution (we refer to this directly above as the success indicator). Although some software tasks may undoubtedly involve more planning than others (e.g., requirements gathering, prioritizing, and scheduling versus coding or application development), others may involve more creativity and intellective faculties (e.g., user experience-related tasks and coding versus black box testing). Similarly, other forms of task may involve more decision-making, cognitive conflict and dilemmas (e.g., requirements gathering and decomposition versus white box testing). These activities, regard- less of their nature, all comprise people processes, which are likely to have a major effect on task performance. Issues such as team trust and culture [136–139], team social motivation [138–140], team norms [141], and moods [134] may interact with group structure [81], the importance of the software task [37] and other interaction-related factors (e.g., information diversity [82,83]) to affect speed of task resolution. We look to examine some of these issues in our subsequent work. We next discuss the implications of our findings in this work for practice and research.

# 7. IMPLICATIONS

The outcomes of this work may support software project governance and software engineering research in a number of ways. For instance, our findings regarding the large numbers of defects and greater member participation in addressing this form of task clearly imply that reducing the incidence of defects would free up additional time for developing and delivering new features. Our results further suggest that a wider spread of practitioners' involvement in requirements gathering may also lessen the incidence of defects, presuming that such issues resulted due to lack of awareness. In addition, approaches that seek to reduce the effort spent on bug fixing, such as techniques for finding duplicate bug reports in repositories or other mechanisms that eliminate or lessen the need for such activities in the first instance (such as TDD and continuous integration practices), may also be helpful. Stakeholders championing software project governance may use these findings to encourage the adoption of such practices during the implementation of their software projects.

Our findings noting the higher level of within-project communication when members were developing new features and extending software functionalities suggest that project participants should be prepared for such increases, and those responsible for project governance should facilitate these communication surges by ensuring the availability of adequate communication channels. Of course, a strategy that promotes increased participation in upfront activities (highlighted above) is likely to shift this emphasis somewhat. Thus, those charged with governing software projects should be vigilant and prepared to make appropriate adjustments should the need arise, in line with the strategy that is adopted.

In fact, beyond such interventions, consideration should also be given to the way practitioners collaborate to perform software development activities. If new members are added to teams that assembled to perform previous work, these new practitioners are likely to need context, and thus, there is likely to be more questions asked and information exchanged to facilitate knowledge transfer and cross-fertilization. Accordingly, increases in these forms of communication should be accommodated. Perhaps scaffolding could also be encouraged to address repetitive questions, as this may reduce the need for ongoing team engagement, and thus, the availability of communication channels. Similarly, knowledgeable software developers should be prepared to express their thoughts and provide contexts for those who are less aware, even when unprompted. This may be particularly necessary for enhancing innovativeness and critical evaluation, which should aid in the delivery of creative solutions. Of course, notwithstanding improvements in task outputs, increases in idea exchange and team members' participation (and related behavioral issues) will likely extend the time taken to complete tasks, and thus, this should be anticipated by project leaders.

In particular, behavioral issues may interact with other task properties, and the more explicit group structure, situational and interaction factors, to affect team performance when software practitioners are performing various forms of software development activities. We look to examine this issue in our future work. Specifically, our follow-up study will seek to examine how differences in team climate and behavioral concerns influence the speed of task resolution. We also intend to examine these issues from project phase to project phase, and for those with different feature portfolios (e.g., user experience versus coding-intensive tasks). We plan to study these issues using various forms of CA techniques. Furthermore, we also plan to examine how various elements of task perception affect the speed with which different software tasks are completed. Other researchers are encouraged to contribute to this knowledge base by conducting similar replication studies.

# 8. LIMITATIONS

Although we have provided a number of insights into this work, we acknowledge that there are several shortcomings that may potentially affect the generalizability of our study outcomes; we consider the following in turn:

(1) **Communication** was assessed based on messages sent around software tasks. These messages were extracted from Jazz, and thus, may not represent all of the project teams' communications. Offsetting this concern is the fact that, as Jazz was developed as a globally distributed project, developers were required to use messages so that all other contributors, irrespective of their physical location, were aware of product and process decisions regarding each WI [100].

(2) Our **directed CA** involving interpretation of textual data is subjective, and hence questions may naturally arise regarding the validity and reliability of the outcomes of this analysis. In this work, we used multiple strategies to mitigate these issues. First, our protocol was adapted from those previously used and tested in the study of interaction and knowledge sharing [108,109], and thus there is a strong theoretical basis for its use. Second, we piloted the protocol and extended our instrument by deriving additional codes directly from the Jazz data, and we tested this extended protocol for accuracy, precision, and objectivity, receiving an interrater measure indicative of excellent agreement [110]. However, we also recognize that additional research using interview techniques would serve to further validate the outcomes of these analyses. In fact, our outcomes in this study do not actually encompass insights into practitioners' perceptions about different attributes of soft- ware tasks. Although our work focused to make inferences largely from the artifacts software practitioners create, we are aware that insights into practitioners' perception would usefully complement and extend our findings.

(3) **Cultural differences** and distance (geographical and temporal) may directly affect software development teams' performance [94], and these phenomena may also have an effect on team members' behaviors – which in turn may lead to performance issues [95]. However, research examining the effects of cultural differences in global software teams has found few cultural gaps and behavioral differences among software practitioners from, and operating in, Western cultures [94]. Given that the practitioners studied in this work all operated in Western cultures, this issue may have had little effect on the results that were observed. In addition, research examining the communication processes of these Jazz practitioners did not find distance to influence the speed of task resolution [37].

(4) Finally, we studied artifacts from **a single organization** using particular development practices. Work processes and work culture at IBM are likely to be specific to that organization and may not be representative of organization dynamics else- where, and particularly for environments that use conventional waterfall processes [1]. Such environments may use more rigid project management practices, with much clearer hierarchical structures, development boundaries, and defined roles [142–144]. Having said that, Costa et al. [145] confirmed that practitioners in the Jazz project exhibited similar coordination needs to practitioners of four projects operating in two distinct companies. Thus, we believe that our results may be applicable to similar large-scale distributed projects.

## 9. CONCLUSION

With the shift of focus from software product characteristics to more team-based issues, and the increased attention given to people and their work practices during software development, repositories have played an increasingly important role in providing artifacts to enable various explorations. Thus far, these explorations have tended to examine various communication metrics, and their linkages to team outcomes. In particular, there has been a wealth of studies that have examined how teams' communication processes relate to the incidence of software bugs. Although these works have provided insights into the way software teams work, given the complexity of human behavior, and different forms of task, we believe that this pool of evidence would also benefit from explorations focused on the wider ecosystem of software development activities, beyond just the study of bugs. Previous studies on group work and task differences have indeed established that there is relevance in understanding how teams work when they are involved in different types of tasks. We anticipated that exploring the way developers work across the range of software tasks, and particularly those initial events that lead to the development of software features in the first place, and then subsequent bugs, could provide added value for the software development community, both in terms of informing team composition strategies and for extending the knowledge in this space.

Accordingly, we have used seminal works on task differences as a theoretical basis for studying how practitioners work when undertaking a range of software tasks. Among our results, we observed that the practitioners studied were mostly involved with fixing defects, and these members engaged the most around new features and software extensions. In addition, software practitioners' engagements covaried depending on the nature of work they were performing. Furthermore, Jazz practitioners took the longest amount of time to deliver new features and software enhancements, and a number of external factors had moderate effects on speed of task resolution. However, given the highly complex nature of software development activities, we believe that behavioral and intrinsic issues may interact with extrinsic variables becoming more significant predictors of the speed of software task resolution. Our initial contextual analysis indeed shows promise and provides initial support for this position. We therefore look to study this issue in future work.

**Sherlock Licorish** is a lecturer in the Department of Information Science at University of Otago, New Zealand. He was awarded a BSc degree from the University of Guyana and MSc and PhD degrees from the Auckland University of Technology (AUT). His research centers on software development process modeling and assessment, the development and provision of software tools, and empirical software engineering and analytics. His analytics work includes topics related to global software development, open source software (OSS) development, and virtual communities. Sherlock's research involves the use of data mining, data visualization, statistical analysis, and other quantitative methods (e.g., social network analysis, linguistic and sentiment analysis, natural language processing (NLP), and probabilistic modeling techniques). He has also used qualitative methods in his research, including qualitative forms of content analysis and dilemma analysis. These techniques (both quantitative and qualitative) are often applied to large repositories and software artefacts. Sherlock is a program committee member for ACSC2016, ACSC2017, EASE2016 and EASE2017, and he reviews articles for IEEE's TSE, I&M, JSS, I&ST, HICSS, AAAI, AMCIS, ECIS and ACIS.

**Stephen MacDonell** is professor of software engineering and director of the Software Engineering Research Laboratory (SERL) at the Auckland University of Technology (AUT) and professor in Information Science at the University of Otago, both in New Zealand. Stephen was awarded BCom (Hons) and MCom degrees from the University of Otago and a PhD from the University of Cambridge. He undertakes research on software analytics and visualization, project planning, estimation and management, information systems development, software forensics, and the application of a range of methods of analysis to software engineering data sets. He is a member of the IEEE Computer Society and the ACM, and serves on the Editorial Board of Information and Software Technology.